\documentclass[a4paper]{jpconf}

\usepackage{graphicx}

\usepackage{wrapfig}

\begin{document}

\title{Modeling neutrino-nucleus interactions in the few-GeV regime.}

\author{N.~Jachowicz, P.~Vancraeyveld}

\address{Ghent University, Department of Physics and Astronomy,

  Proeftuinstraat 86, B-9000 Gent, Belgium.}

\ead{natalie.jachowicz@ugent.be}

\begin{abstract}

Detecting neutrinos and extracting the information they bring along is
an ambitions task that requires a detailed understanding of neutrino-nucleus
interactions over a broad energy range. 
We present calculations for quasi-elastic neutrino-induced nucleon
knockout reactions on atomic nuclei and neutrino-induced pion
production reactions.
In our models, final-state interactions are introduced using a relativistic
multiple-scattering Glauber approximation (RMSGA) approach.  For
interactions at low incoming neutrino energies, long-range
correlations are implemented by means of a continuum random phase
approximation (CRPA) approach.
As neutrinos are the only particles interacting solely by means of
the weak interaction, they can reveal information about e.g. the
structure of nuclei or the strange quark content of the nucleon that
is difficult to obtain otherwise.  We investigated these effects and
present results for  the sensitivity of neutrino interactions to
the influence of the nucleon's strange quark sea.

Contribution to NUFACT 11, XIIIth International Workshop on Neutrino
Factories, Super beams and Beta beams, 1-6 August 2011, CERN and
University of Geneva (submitted to IOP conferende series). 

\end{abstract}

\section{Introduction}

The  experimental confirmation of  neutrino oscillations raised an enormous experimental and theoretical interest in the oscillation properties of these particles.
For an accurate interpretation of the experimental data, a thorough
understanding of the interactions involved in oscillation experiments
is indispensable.  Cross sections for neutrino-nucleus interactions
play an important role.  Whereas the experimental observable  is often
an inclusive
cross section, a detailed study of exclusive cross sections is indispensable
for a thorough understanding of these processes.

\section{Quasi-elastic neutrino scattering}\label{qe}  

 The one-fold differential cross section for the scattering process $A(\nu,lN)$ is given by
\begin{equation}
\frac{d\sigma}{dT_N}=\frac{M_N M_{A-1}}{(2\pi)^3 M_A} 4\pi^2 \int \sin\theta_ld\theta_l\int\sin\theta_Nd\theta_N k_N f_{rec}^{-1}\sigma_M\left[v_LR_L+v_TR_T+hv_{T'}R_{T'}\right],\label{cs1}
\end{equation}
with $M_N$, $T_N$ and $\vec{k}_N$ the mass, kinetic energy and momentum of the ejectile, $M_A$ and $M_{A-1}$ the mass of the target and residual nuclei.  The direction of the outgoing  lepton and nucleon is determined by $\Omega_l(\theta_l,\phi_l)$ and $\Omega_N(\theta_N,\phi_N)$. The recoil factor is denoted by $f_{rec}$. The quantity $\sigma_M$ is the weak variant of the Mott cross section.
   \begin{figure*}[tb]
\vspace*{7.cm}
\special{hscale=38 vscale=38 hsize=1500 vsize=600
         hoffset=60 voffset=-22 angle=-0 psfile="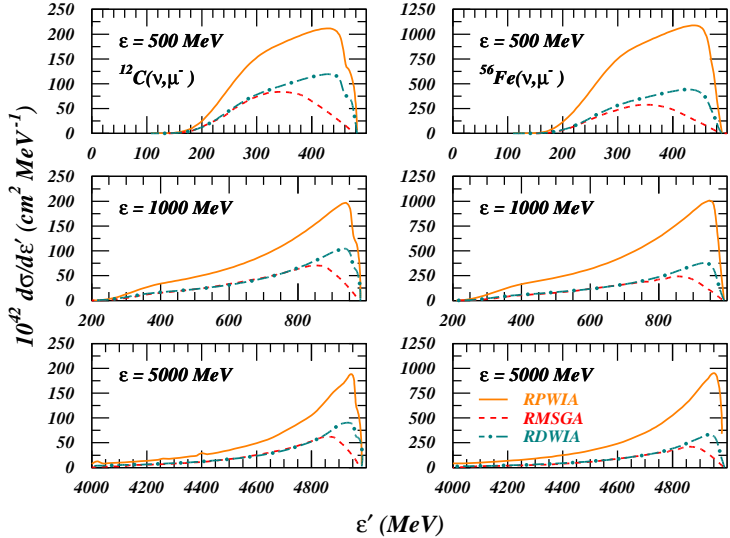"}
\caption{Cross sections for the charged-current reactions $^{12}$C$(\nu_{\mu}, \mu^-)$ and $^{56}$Fe$(\nu_{\mu}, \mu^-)$ as a function of the energy of the outgoing muon $\varepsilon'$.  The different curves compare the RMSGA results (dashed) with the RDWIA (dash-dotted) and the RPWIA limit (full line).}
\label{figfsi}
\end{figure*}
In Eq.(~\ref{cs1}), $v_L$, $v_T$ and $v_{T'}$ are the longitudinal, transverse and interference kinematic factors and $R_L$, $R_T$ and $R_{T'}$ the accompanying structure functions, reflecting the influence of nuclear dynamics on the scattering process \cite{us}. The helicity of the incoming neutrino is denoted by $h$.
  In our numerical calculations, bound-state wave-functions are obtained within the Hartree approximation to the $\sigma-\omega$ model, adopting the W1 parametrization for the different field strengths.

We introduce final-state interactions adopting a
relativistic multiple-scattering Glauber approximation (RMSGA)
\cite{us,dimi}. As a semi-classical approach, this technique exploits
the advantages of the kinematics conditions reigning at sufficiently
high energies, where high momentum transfers strongly favor forward
 rescattering of the outgoing nucleon.  The Glauber technique
assumes linear trajectories for the ejectile and frozen spectator
nucleons in the residual system.  The influence of the nuclear medium
on the outgoing nucleon's wave function are condensed in the eikonal
phase ${\cal G}[\vec{b}(x,y),z]$, that summarizes the effects of the
scattering reactions the ejectile undergoes.  This results in a
scattering wave function that can be written as ${\phi}_F(\vec{r})={\cal G}[\vec{b}(x,y),z]\;\phi_{k_N,s_N}(\vec{r})$,
with $\phi_{k_N,s_N}(\vec{r})$ a relativistic plane wave.  In the
limit of vanishing final-state interactions (${\cal G}=1$) , the
formalism becomes equivalent to the relativistic plane wave impulse
approximation (RPWIA) \cite{us}.
Figure~\ref{figfsi} shows the influence of final state interactions on
cross sections for charged-current processes on $^{12}$C and
$^{56}$Fe, and compares Glauber and RDWIA calculations.  In the region
where both approaches are valid, their results are in excellent agreement.

\section{Strangeness}

\begin{figure}[tb]
\vspace*{4cm}
  \begin{center}
\special{hscale=60 vscale=60 hsize=1500 vsize=600
         hoffset=50 voffset=-10 angle=0 psfile="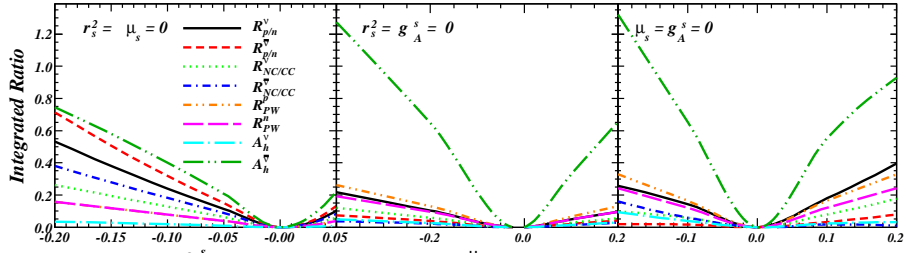"}
   \caption{Comparison between the strangeness influence on various ratios of total cross-sections in terms of the relative sensitivity $\left|\frac{R(s=0)-R(s)}{R(s=0)}\right|$, as a function of the strangeness form factors $g_A^s$, $\mu_s$ and $ r_s^2$ for 1 GeV neutrino scattering off $^{12}$C.}
\label{vergelijkingRatio}
  \end{center} 
\end{figure}

The strangeness content of the nucleon influences neutrino-nucleus
cross sections and has an important impact on several cross-section ratios.
We compared the influence of axial as well as vector strangeness on
$\nu$ and $\overline{\nu}$ cross-section ratios \cite{strang}.
We compared  the impact of the weak strangeness form factors on the ratio of proton-to-neutron knockout, the ratio of neutral-to-charged current cross sections, on the Paschos-Wolfenstein relation
for protons and neutrons, and on the longitudinal helicity asymmetry for neutrinos and antineutrinos.
Figure~\ref{vergelijkingRatio} summarizes the main results.
The longitudinal helicity asymmetry for antineutrinos is most sensitive to strangeness effects.  In general, antineutrino-induced processes exhibit a more outspoken strangeness sensitivity than their neutrino counterparts.
The overall sensitivity of $R_{NC/CC}$ ratios to strangeness effects
is considerably smaller than that of $R_{p/n}$.

Whereas in PVES the tininess of the axial strangeness effects  impedes the determination of $g_A^s$, in neutrino scattering a thorough understanding of vector strangeness effects is a prerequisite for extracting information on the axial strangeness. Hence a combined  analysis of  parity-violating electron scattering and neutrino-induced processes would  offer the best prospects for a thorough understanding of the influence of the nucleon's strange quark sea on electroweak processes.

\section{Weak one-pion production on a nucleus}

\begin{wrapfigure}[13]{l}{0.7\textwidth}
\vspace{3.9cm}
\special{hscale=32 vscale=32 hsize=1500 vsize=600
         hoffset=25 voffset=-17 angle=-0 psfile="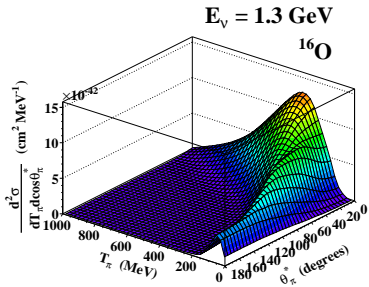"}
\caption{Resonant pion production off $^{16}$O for 1.3 GeV incoming
  neutrinos.}
\label{respi}
\end{wrapfigure}

At slightly higher energies, resonant  and coherent pion production
become important reaction channels. The $\Delta$-mediated pion-pro\-duc\-tion cross section is evaluated adopting the relativistic plane-wave
impulse approximation for the calculation of the matrix element.  The
single particle wave functions are the same as those used in
section \ref{qe}.  Medium modifications for the $\Delta$ were
implemented as discussed in Ref.~\cite{chris}.  Figure \ref{respi} shows
the differential cross section for neutrino-induced pion-production
off $^{16}$O.

\begin{wrapfigure}[9]{r}{0.75\textwidth}
\vspace{2.7cm}
\special{hscale=25 vscale=25 hsize=1500 vsize=600
         hoffset=80 voffset=-10 angle=-0 psfile="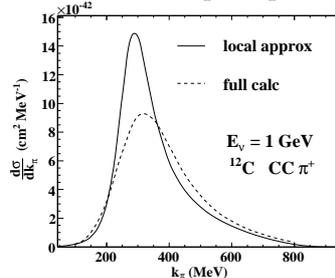"}
\caption{Comparison of full and local approximation calculations for coherent pion production.}
\label{cohpi}
\end{wrapfigure}

Coherent pion production leaves the nuclear target intact by
transferring only small four-momenta.  Figure \ref{cohpi} compares
full- and local-approximation calculations for charged-current
coherent pion production on $^{12}$C.

\section{Long-range CRPA correlations}

\begin{wrapfigure}[18]{l}{0.65\textwidth}
\vspace*{4.7cm}
\special{hscale=60 vscale=60 hsize=1500 vsize=600
         hoffset=25 voffset=-35 angle=-0 psfile="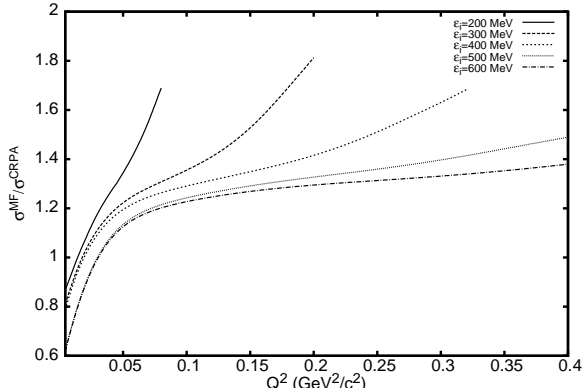"}
\caption{Ratio of cross sections obtained with mean field (MF)
   wave functions to cross sections including 
  continuum RPA correlations (CRPA) as a function of $Q^2$ for incoming
  neutrino-energies ranging from 200 to 600 MeV and with $^{12}$C as
  target nucleus.}
\label{fig1}
\end{wrapfigure}

For inclusive cross sections, especially at higher energies, the Fermi
Gas model is remarkably accurate.  At lower energies however, nuclear
effects become preponderant.
In a Random Phase Approximation (RPA) approach,
long-range correlations between the nucleons in the nucleus are
introduced.
Whereas in a mean-field calculation a nucleon experiences the presence
of the others only through the mean-field generated by their mutual
interaction, the random phase approximation  additionally allows
the particles to interact by means of  the residual two-body force.  
  The random phase approximation describes a
nuclear state as the coherent superposition of particle-hole
contributions  \cite{crpaus1,crpaus2}.
\begin{equation}
\left| \Psi_{RPA} \right\rangle = \sum_c \, \left\{\, X_{(\Psi,C)} \;\left| ph^{-1}\right\rangle\: -\: Y_{(\Psi,C)} \;\left| hp^{-1}\right\rangle \, \right\} .\label{xyrpa}
\end{equation}
The summation index $C$ stands for all quantum numbers
defining a reaction channel unambiguously.
In our model, the continuum RPA equations are solved using a Greens function approach in
which the polarization propagator is approximated by an iteration of
the first-order  contribution. The unperturbed
wave functions are generated using either a Woods-Saxon potential or a
HF-calculation using a Skyrme force.  The latter approach makes
self-consistent HF-RPA calculations possible. 
As is shown in
figure~\ref{fig1}, long-range Random
Phase Approximation  correlations  account for a considerable reduction of cross
sections at low incoming neutrino-energy.

\end{document}